\documentclass[12pt]{article}

\usepackage{epsfig}

\usepackage{amssymb}
\usepackage{amsfonts}

\usepackage{color}
 
\def\be{\begin{equation}}
\def\ee{\end{equation}}
\def\ba{\begin{array}{c}}
\def\ea{\end{array}}

\newcommand{\bea}{\begin{eqnarray}}
\newcommand{\eea}{\end{eqnarray}}

\newcommand{\kt}{\rangle}
\newcommand{\br}{\langle}

\begin{document}

. \vspace{0.1cm}

 \begin{center}{\Large \bf

Reconstruction of full-space quantum Hamiltonian from its
effective, energy-dependent model-space projection

  }

\vspace{1cm}

  {\bf Miloslav Znojil}$^{a,b,c}$

\end{center}

\vspace{3mm}

 $^{a}${The Czech Academy of Sciences,
 Nuclear Physics Institute, 
 Hlavn\'{\i} 130,
250 68 \v{R}e\v{z}, Czech Republic\footnote{{e-mail:
znojil@ujf.cas.cz}}}


 $^{b}${Department of Physics, Faculty of
Science, University of Hradec Kr\'{a}lov\'{e}, Rokitansk\'{e}ho 62,
50003 Hradec Kr\'{a}lov\'{e},
 Czech Republic}

 $^{c}${Institute of System Science, Durban University of Technology,
Durban, South Africa}

\vspace{5mm}


\section*{Abstract}

An unknown and to-be-reconstructed
full-space quantum Hamiltonian $H$
is assumed to
have an $N$ by $N$ matrix form
which is,
out of a subspace
of Hilbert space, tridiagonal.
The reconstruction of $H$
from a given, energy-dependent
effective model-space Hamiltonian
$H^{}_{ef\!f}(E)$
is then shown feasible.
The process consists of two steps.
In the first one the
information
about dynamics available via
$H^{}_{ef\!f}(E)$
is rearranged
using matrix inversion.
In a way illustrated via a few simplest
special cases,
a well-defined coupled set of
polynomial algebraic equations has
to be then solved in
the second step.

\subsection*{Keywords}

.

approximations using ``model-space'' subspaces of Hilbert space;

energy-dependent phenomenological quantum models of bound states;

suppression of the energy-dependence of effective Hamiltonians;

exact Hamiltonian reconstructed from its model-space projection;

%
%

\newpage

\section{Introduction\label{I}}

Quantum theory acquired its name, cca 100 year ago, after the
Heisenberg's proposal \cite{Heisenberg}
that the ``quantized'' energy spectra as observed
in certain microscopic systems
could prove
tractable as eigenvalues of some suitable $N$ by $N$
matrix Hamiltonians
with $N \leq \infty$
\cite{Messiah}.
In the related $N-$dimensional Hilbert space ${\cal H}$
the
stationary bound states of
the system can be then specified
as solutions of a linear algebraic eigenvalue problem
 \be
 H_{}\,|\psi^{(n)}\kt = E^{(n)}\,|\psi^{(n)}\kt
 \,,\ \ \ \
  n =  1, 2,\,\ldots
 \,.
 \label{bsthee}
 \ee
In applications, nevertheless,
the solution of Eq.~(\ref{bsthee})
need not be easy, especially
when one decides to study a
not too elementary quantum system
like, say, a larger molecule or a heavier atomic
nucleus.

More than 60 years ago, Feshbach \cite{Feshbach}
and L\"{o}wdin \cite{Loewdin}
addressed the problem and
made an important observation that
not all of the matrix elements
$H_{ij}$ contribute to the determination of
the first few energy levels $E^{(n)}$ with the same weight.
This weight
will vary, of course, with the changes of the basis
in Hilbert space.
Still, one often observes that
such a weight of elements $H_{ij}$  may quickly
decrease with
the growth of the subscripts $i$ and/or $j$.
The Hamiltonian matrix may be then
partitioned,
 \be
 H_{}=
  \left[ \begin {array}{ccc|ccc}
  H_{11}&\ldots&H_{1M}&H_{1M+1}&\ldots&H_{1N}\\
  \vdots&\ddots&
 \vdots&\vdots&\ddots&\vdots
    \\
  H_{M1}&\ldots&H_{MM}&H_{MM+1}&\ldots&H_{MN}\\
  \hline
 H_{M+11}&\ldots&H_{M+1M}&H_{M+1M+1}&\ldots&H_{M+1N}\\
  \vdots&\ddots&
 \vdots&\vdots&\ddots&\vdots
    \\
 H_{N1}&\ldots&H_{NM}&H_{NM+1}&\ldots&H_{NN}\\
 \end {array} \right]\,
 \label{pasrtg}
 \ee
with the relevance of its elements dominated
by its upper left ``model-space''
submatrix.

In such a setting
people usually recall the variational philosophy and employ
a brute-force $M$ by $M$ truncation of the matrix.
Immediately,
this can lead to a conflict
between the speed and precision of the
calculations. As indicated above,
one of the most consequent resolutions of such a conflict
has been found
by Feshbach \cite{Feshbach} and/or, independently,
by L\"{o}wdin~\cite{Loewdin}.
They both proposed
to proceed in two steps. In the first step one
replaces the exact and
complete $N-$dimensional physical Hilbert space ${\cal H}$
by its smaller, $M-$dimensional subspace
called, usually, model space.
In a way which will be briefly outlined in section \ref{duha} below,
such a reduction of space has been accompanied, in the second step, by
a rigorous non-numerical replacement
of the full-fledged realistic Hamiltonian (\ref{pasrtg})
by its
formally isospectral
``effective''
$M$ by $M$ matrix alternative which is
to be denoted, in what follows, by a dedicated symbol
$H^{}_{ef\!f}(E)$
indicating that the latter matrix must be, by definition,
energy-dependent.

The
energy-dependence
of $H^{}_{ef\!f}(E)$
makes the model-space eigenvalue problem
 \be
 H_{ef\!f}(\eta)\,|\phi^{(n)}\kt = E^{(n)}\,|\phi^{(n)}\kt
 \,,\ \ \ \ \eta=E^{(n)}
 \,,
 \ \ \ \ \ n =  1, 2,\,\ldots \,
 \label{ethee}
 \ee
nonlinear: After
an initial trial-and-error
tentative choice of $\eta$ and of a fixed index $n=n_0$
we solve Eq.~(\ref{ethee}) but we only
get a tentative spectrum of energies $E^{(n)}_{initial}=E^{(n)}(\eta)$.
It remains unphysical
whenever $\eta \neq E^{(n_0)}$.
We have to amend the choice of $\eta$. This means that in practice
we have to iterate the process
until we get a consistent solution such that
$E^{(n_0)}_{f\!inal}=E^{(n_0)}=E^{(n_0)}(E^{(n_0)})$.

This is a mathematical difficulty
which may slow down the calculations.
The strength and the influence
of the nonlinearity
could only be weakened via a transition to a larger
model space.
Obviously, this would necessitate
an amendment of
the effective Hamiltonian
$H^{}_{ef\!f}(E)=H^{(M)}_{ef\!f}(E)$
during a step-by-step increase
of the model space dimension
$M \to M+1 \to M+2 \to \ldots$.
This is precisely the problem which is to be addressed in our present paper.

It seems that
such a form of amendment
has not been considered in the literature.
At the same time,
whenever one would find such a reconstruction feasible,
its practical consequences
would be important in multiple areas of applied quantum physics.
In this sense, our present paper can be read as
the first step towards this goal.
The overall structure of the process of reconstruction
will be outlined in sections
\ref{oforof} and
\ref{orof}
while
a few samples of an explicit implementation
of the recipe will be given in section
\ref{inpro}.
In the context of possible physical applications,
finally, our results will be
discussed in section
\ref{discussion}
and summarized in section \ref{summary}.

\section{Quantum Hamiltonians reduced to a subspace\label{duha}}

The
determination of
properties of many quantum systems
is often dominated by not too many matrix
elements of the respective
operators of observables.
For example, when we need to determine the
low-lying bound-state energies of
a system described by a complicated
Hamiltonian $H$
we may reveal that a fairly reasonable
precision is already obtainable
by the diagonalization of
a drastically truncated matrix form of $H$.

The success of truncations
depends, critically, on our choice of the basis
in Hilbert space as well as on our specification
of the relevant
model space determined by
projector
 \be
 P=\sum_{m=1}^{M}|m\kt \br m|
 \,.
 \label{enproj}
 \ee
Several robust
strategies of calculations can be then used in practice.
One of the most systematic ones
is based on the Feshbach-inspired
construction of corrections
reflecting the influence of the $Q-$projected subspace
with $Q=I-P$.

\subsection{The Feshbach's concept of effective Hamiltonian $H^{}_{ef\!f}(E)$}

The inclusion of corrections makes
the effective bound-state
Hamiltonian energy-dependent in general (cf. Eq.~(\ref{ethee}) above).
In principle, this would even open the possibility of
an exact isospectrality of  $H^{}_{ef\!f}(E)$
with the full-fledged $H$.

The latter goal can be achieved via
partitioning (\ref{pasrtg}) which can be rewritten in compact form,
 \be
 H_{}= \left[ \begin {array}{cc}
  PH_{}P&PH_{}Q\\
  QH_{}P&QH_{}  Q
 \end {array} \right]
 \,.
 \label{pafukit}
 \ee
After
a variationally motivated choice of a sufficiently large
model-space dimension $M$, even the brute-force
truncation $ H \to PHP$ of the Hamiltonian
may lead to a useful
approximate result.
Still, when we have to solve a realistic Eq.~(\ref{bsthee})
we are often surprised by the contrast between
the easiness of the preliminary truncation-dependent estimates
and the emergence of
nontrivial difficulties when one needs and tries to
improve the precision \cite{Dyson}.

As we already indicated,
one of the standard mathematical tools
of achieving this goal
is offered by
effective
Hamiltonian. Its formal definition
is based on the partitioning
of
the full-space eigenvalue problem,
 \be
 \left (
 \begin{array}{cc}
 P(H_{}-E^{(n)})P&PH_{}Q\\
 QH_{}P&Q(H_{}-E^{(n)})Q
 \end{array}
 \right )\,
 \left (
 \begin{array}{cc}
 P\,|\psi^{(n)}\kt\\
 Q\,|\psi^{(n)}\kt
 \end{array}
 \right )=0\,,
 \ \ \ \ n = 1, 2, \ldots
 \,.
 \label{[u2]}
 \ee
Once we intend to get
an efficient amendment
of the rather naive brute-force truncation,
we may decide to treat
the influence of the
out-of-the-model-space submatrices
$PHQ$, $QHP$ and $QHQ$
in a nontrivial manner.

The
components of the wave functions which lie
out of the $P-$projected model space
may be expected to
play a less essential role.
We may formally
eliminate these
components.
Using
the full-space
Schr\"{o}dinger equation~(\ref{[u2]})
we get their explicit definition
in terms of their model-space-projection complement
$|\phi^{(n)}\kt=
 P\,|\psi^{(n)}\kt$,
 \be
 Q\,|\psi^{(n)}\kt=
 -Q\,\frac{1}{QH_{} Q-E^{(n)}}\,Q\,H_{}\,|\phi^{(n)}\kt
 \,.
 \label{[3]}
 \ee
A routine backward insertion of the latter formula
in the first line of Eq.~(\ref{[u2]})
yields
the amended, model-space-restricted
new Schr\"{o}dinger equation (\ref{ethee})
which contains the effective Hamiltonian
 \be
 H_{\!\it ef\!f}(E)=P \,H_{}\,P
 + P\,H_{}\,Q\,\frac{Q}{E-Q\,H_{}\,Q}\,Q\,H_{}\,P
 \,.
 \label{formu}
 \ee
Due to the energy-dependence of $ H_{\!\it ef\!f}(E)$,
both of the alternative formulations (\ref{bsthee}) and (\ref{ethee}) of
Schr\"{o}dinger equation remain strictly formally equivalent.

\subsection{Method of construction}

The direct process of the construction
of $H_{\!\it ef\!f}(E)$
from the full-space Hamiltonian $H$
requires the inversion of
matrix $Q\,(H-E)\,Q$
(cf. Eq.~(\ref{formu})).
In general, this is a rather
nontrivial numerical task.
Fortunately, its feasibility can be
perceivably facilitated by
a unitary transformation which would make
the out-of-the-model-space
submatrix of the Hamiltonian tridiagonal,
 \be
 Q H Q=
  \left[ \begin {array}{ccccc}
 a_1&b_1&0
 &\ldots&0
   \\
  c_2&a_2&b_2&\ddots
 &\vdots
   \\
 0
 &\ddots&\ddots&\ddots&0
   \\
 \vdots&\ddots&c_{K-1}&a_{K-1}&b_{K-1}
    \\
  0&\ldots&0&c_{K}&a_{K}
    \\
 \end {array} \right]\,.
 \label{eitik}
 \ee
The inversion
can be then facilitated using
factorization ansatz
 \be
 Q\,(H-E)\,Q=
 {\cal U}\,
 {\cal F}\,
 {\cal L}\,
 \label{[9]}
 \ee
in which
the central matrix factor ${\cal F}$ is diagonal, with elements
 $$
 1/f_1,  1/f_2, \ldots,  1/f_K\,.
 $$
We find that
for Hamiltonians (\ref{eitik}) the factorization (\ref{[9]})
with
 \be
 {\cal U}=
  \left[ \begin {array}{ccccc}
  1&b_1f_2&0
 &\ldots&0
   \\
     0&1&b_2f_3&\ddots
 &\vdots
   \\
 0
 &0&\ddots&\ddots&0
   \\
 \vdots&\ddots&\ddots&1&b_{K-1}f_K
    \\
  0&\ldots&0&0&1
    \\
 \end {array} \right]\,,
 \ \ \ \ \ \
 {\cal L}=
  \left[ \begin {array}{ccccc}
  1&0&0
 &\ldots&0
   \\
     f_2c_2&1&0
 &\ldots&0
   \\
   0&f_3c_3&\ddots&\ddots
 &\vdots
   \\
 \vdots&\ddots
 &\ddots&1&0
   \\
  0&\ldots&0&f_Kc_{K}&1
    \\
 \end {array} \right]\,
 \label{lowkit}
 \ee
is an algebraic identity if and only if
 \be
 f_k=\frac{1}{a_k-E-b_kf_{k+1}c_{k+1}}\,,\ \ \ \
 k=K, K-1,\ldots,2 ,1 
 \label{cf}
 \ee
where we set $f_{K+1}=0$.

In the definition
of the Feshbach's effective Hamiltonian (\ref{formu}) the
factorization (\ref{[9]}) implies
the closed-form factorization of the
out-of-the-model-space
resolvent,
 \be
 \frac{Q}
 {Q\,(H-E)\,Q}=
 {\cal L}^{-1}\,
 {\cal F}^{-1}\,
 {\cal U}^{-1}\,.
 \label{[9h]}
 \ee
Once we abbreviate
$\alpha_{k+1}=-b_kf_{k+1}$ and $\beta_j=-c_jf_j$
we get the explicit formulae
 \be
 {\cal U}^{-1}=
  \left[ \begin {array}{ccccc}
  1&\alpha_2&\alpha_2\alpha_3
 &\ldots&\alpha_2\alpha_3\ldots\alpha_K
   \\
     0&1&\alpha_3&\ddots
 &\vdots
   \\
 0
 &0&\ddots&\ddots&\alpha_{K-1}\alpha_K
   \\
 \vdots&\ddots&\ddots&1&\alpha_K
    \\
  0&\ldots&0&0&1
    \\
 \end {array} \right]\,,
 \ \ \ \ \ \
 {\cal L}^{-1}=
  \left[ \begin {array}{ccccc}
  1&0&0
 &\ldots&0
   \\
     \beta_2&1&0
 &\ldots&0
   \\
   \beta_3\beta_2&\beta_3&\ddots&\ddots
 &\vdots
   \\
 \vdots&\ddots
 &\ddots&1&0
   \\
   \beta_K \ldots \beta_3\beta_2&\ldots&\beta_K\beta_{K-1}&\beta_K&1
    \\
 \end {array} \right]\,.
 \label{inlowkit}
 \ee
This explains why it makes sense to represent
any exact, full-space Hamiltonian of interest in the
special basis in which it acquires
the partially tridiagonalized
matrix form as specified by Eq.~(\ref{eitik}).
The efficiency
and usefulness of the construction of the
model-space
$M$-by-$M$-matrix projection $H_{\!\it ef\!f}(E)$
of the full-space $N$ by $N$ Hamiltonian $H$
appear strongly enhanced by such a specific choice
of the basis.

\section{Problem of reconstruction\label{oforof}}

In the preceding paragraph we saw that
even in the ``direct'' problem of
transition from $H$ to $H_{\!\it ef\!f}(E)$
the related technicalities
become significantly facilitated
under assumption (\ref{eitik}).
This observation encouraged us
to accept the same
partial-tridiagonality
postulate also in what follows.
We revealed that
a comparatively much more difficult
``inverse'' problem concerning the
reconstruction of $H$ from $H_{\!\it ef\!f}(E)$
can be found tractable under such a constraint.

\subsection{Motivation: Two types of the energy-dependence of $H_{\!\it
ef\!f}(E)$}

During a finite-precision
calculation of a few low-lying bound-state energies
one has to distinguish between
the following two alternative scenarios:

\begin{description}

\item{[a]}
the energy dependence of the effective Hamiltonian
remains negligible. Without any significant loss of precision
one may then
choose and fix a more or less arbitrary
suitable parameter $\eta$ and
solve Eq.~(\ref{ethee})
using a formally
energy-independent approximate operator
$H_{\!\it ef\!f}(\eta)$.

\item{[b]}
the effective Hamiltonian
is perceivably energy-dependent so that in general,
operator
$H_{\!\it ef\!f}(\eta_1)$
differs from
$H_{\!\it ef\!f}(\eta_2)$
when $\eta_1\neq \eta_2$.

\end{description}

 \noindent
In the case of option [a]
the negligibility of the differences between the exact model
$H_{\!\it ef\!f}(E)$ and
the approximate model $H_{\!\it ef\!f}(\eta)$
enables us to work just with the latter
$M$ by $M$ matrix, treating it
simply as a
practically equivalent substitute for the initial
larger matrix. This means that
the validity of
classification [a] offers a
good reason for transition from
the full Hilbert space of dimension $N$
to a ``diminished full space'' of
dimension $M$.
In other words,
just an inessential loss of precision
would be caused by the replacement of the
full-space Schr\"{o}dinger
Eq.~(\ref{bsthee}) by its
smaller-space
alternative~(\ref{ethee}),
especially when we
need to calculate just a
low-lying part of the
bound-state spectrum.

In what follows we will pay attention to
the other, nontrivial scenario [b].
While not knowing the
``exact''
(i.e., energy-independent)
$N$ by $N$ Hamiltonian $H$ of Eq.~(\ref{pasrtg}),
we assume to be given just
its energy-dependent
model-space projection $H_{ef\!f}(E)$.
We are going to demonstrate that
the reconstruction of $H$
from the available
effective Hamiltonian
of class [b]
may still remain,
under certain conditions, feasible.

The
feasibility of such a project appeared to
require several simplifications.
First, we will only consider
such a subset of
the models of class [b]
for which the difference $K=N-M$ is not too large.
This is a requirement which was
partially inspired by the recommended
decrease of
$N$ in the case of
the effective Hamiltonians with property [a].
In the analogous context of scenario [b],
the smallness of the increase
of dimension $M \to M+K$
need not eliminate
the $E-$dependence completely,
but
one can expect that
such an unwanted source of nonlinearity
of Eq.~(\ref{ethee}) becomes
perceivably
suppressed at least.

\subsection{Assumption of a doorway-state-mediated decoupling}

One of our other,
technically equally important simplifications
reflects, once more, the above-mentioned
decreasing relevance
of
the out-of-the-model-space
matrix elements of $H$.
In this sense, a
weakening of the
energy dependence of
$H_{ef\!f}(E)$.
will have to be achieved by a constructive
enlargement of the model space.


For this purpose we propose a decomposition $P=P^{(-)}+D$
of projector $P$
into a pair of sub-projectors.
This induces a
refinement of partitioning (\ref{pafukit}),
 \be
 H=
  \left[ \begin {array}{c|c||c}
  P^{(-)}\,H\,  P^{(-)}\,& P^{(-)}\,H\,{D}&P^{(-)}\,H\,Q
   \\
   \hline
  {D}\,H\,  P^{(-)}\,&{D}\,H\,{D}&{D}\,H\,Q
   \\
   \hline \hline
  Q\,H\,  P^{(-)}\,&Q\,H\,{D}&Q\,H\,Q
 \end {array} \right]\,.
 \label{3sp}
 \ee
Now we can characterize the effective Hamiltonian
living in the
$P-$projected
model space as ``well behaved'', i.e.,
belonging ``almost'' to class [a].
Its smaller,
$P^{(-)}-$projected  model-space partner
can be then assumed
``less-well-behaved'', i.e.,
more wildly energy-dependent
and, hence, belonging,
without any doubts,  to class [b].


It is worth adding that a
similar
sub-partitioning can also be encountered in
the many purely pragmatic applications
of Schr\"{o}dinger equations.
The reason usually lies in the
acceptance of the
so called
``doorway-state-hypothesis''.
By this hypothesis
one is allowed to neglect the
elements $H_{ij}$ in which the {\em difference\,} between $i$ and $j$ is
``sufficiently large''.
In our present notation the doorway-state-hypothesis
may be then given the form
of omission of the coupling
between the $P^{(-)}-$ projected and $Q-$projected
states,
 \be
 P^{(-)}\,H\,Q = 0\,,\ \ \ \ Q\,H\,P^{(-)} = 0\,.
 \label{tydli}
 \ee
Such a form of the
doorway choice of a special
basis in Hilbert space
leads to the
manifestly block-tridiagonal-matrix form of the
Hamiltonian,
 \be
 H=
  \left[ \begin {array}{c|c|c}
  P^{(-)}\,H\,  P^{(-)}\,& P^{(-)}\,H\,{D}&0
   \\
   \hline
  {D}\,H\,  P^{(-)}\,&{D}\,H\,{D}&{D}\,H\,Q
   \\
   \hline
  0&Q\,H\,{D}&Q\,H\,Q
 \end {array} \right]\,.
 \label{doo3}
 \ee
The most important consequence is that also the
effective Hamiltonian itself becomes simplified,
 \be
 H_{\!\it ef\!f}(E)=P \,H\,P + Z(E)\,,
 \ \ \ \
 Z(E) = D\,H\,Q^{}\,
 \frac{Q^{}}{E-Q^{}\,H\,Q^{}}\,Q^{}\,H\,D
 \,.
 \label{wwformu}
 \ee
We will use here just
a minimal (i.e., one-dimensional)
doorway-state subspace, i.e., projectors
 \be
 P^{(-)}=\sum_{m=1}^{M-1}|m\kt \br m|\,,\ \ \ \
 {D}=|M\kt \br M|\,,\ \ \ \
 Q=\sum_{k=1}^{K}|M+k\kt \br M+k|
 \,.
 \label{isu}
 \ee
As long as
$P^{(-)}+{D}+Q=I$ and
 ${M} ={M}^{(-)}+1 \geq 2$
we will have to deal just with the
``to be reconstructed'' Hamiltonian matrices
 \be
 H=
  \left[ \begin {array}{ccc|c|ccc}
  H_{11}&\ldots
 &H_{1M-1}&H_{1M}&0&\ldots&0
   \\
 \vdots&
 &\vdots&\vdots
 & \vdots&& \vdots
    \\
 H_{M-11}&\ldots
 &H_{M-1M-1}&H_{M-1M}&0&\ldots
 &0
    \\
 \hline
 H_{M1}&\ldots
 &H_{MM-1}&H_{MM}&H_{MM+1}&\ldots&
 H_{MM+K}
   \\
 \hline
     0&\ldots
 &0&H_{M+1M}&H_{M+1M+1}&\ldots&
 H_{M+1M+K}
   \\
 \vdots&&\vdots
 &\vdots&\vdots&&\vdots
   \\
  0&\ldots&0&H_{M+KM}&H_{M+KM+1}&\ldots&
 H_{M+KM+K}
    \\
 \end {array} \right]\,.
 \label{oukitie}
 \ee
For such an input
the exact effective Schr\"{o}dinger operator of Eq.~(\ref{ethee})
is $M$ by $M$ matrix
 \be
 H_{\!\it ef\!f}(E)-E=
  \left[ \begin {array}{cccc}
  H_{11}-E&\ldots
 &H_{1M-1}&H_{1M}
   \\
 \vdots&\ddots
 &\vdots&\vdots
    \\
 H_{M-11}&\ldots
 &H_{M-1M-1}-E&H_{M-1M}
    \\
 H_{M1}&\ldots
 &H_{MM-1}&{\cal G}(E)
   \\
 \end {array} \right]\,.
 \label{finkit}
 \ee
where
 \be
 {\cal G}(E)=
 H_{MM}-E
 +\left [H\,Q\,\frac{Q}{E-Q\,H\,Q}\,Q\,H\right ]_{MM}
 \,.
 \label{fundame}
 \ee
This is the statement which follows from the insertion of
Hamiltonian (\ref{oukitie})
in definition (\ref{formu})
or, better,
in its simplified form (\ref{wwformu}).

Matrix (\ref{finkit}) represents
the exact model-space-reduced Schr\"{o}dinger operator
of Eq.~(\ref{ethee})
which differs from its ``unperturbed''
truncated approximation
$P(H-E)P$
in a single (but energy-dependent) matrix element ${\cal G}(E)$.
By construction, this element is
given by expression (\ref{fundame}).
Information about the whole spectrum which
is encoded
in Hamiltonian $H$
can be perceived as carried
by such a non-linear function of energy.

In what follows, precisely such a function ${\cal G}(E)$
(fitting, in principle, some hypothetical experimental data)
will be assumed
given in advance.

\section{The method of reconstruction\label{orof}}

In experimental physics using an {\it ad hoc\,}
effective Hamiltonian
$H_{\!\it ef\!f}(E)$, the measured data
are often fitted
via
Schr\"{o}dinger
Eq.~(\ref{ethee}).
Such a rather pragmatic approach
to quantum theory can prove descriptively
successful
even when it is
not based on the knowledge of $H$ in full space.
We  believe that in such a situation
one of the most consequent ways of restoration
of the credibility of the model
has to be sought
in a return to
definition (\ref{formu})
and in a reconstruction
of the full-space Hamiltonian $H$.
An outline of one of
such model-verification strategies
is to be described now in more detail.

\subsection{Final form of ansatz}

In order to make such a project feasible
we already
restricted our attention to the specific
doorway-state-controlled models (\ref{oukitie}).
In addition,
we will recall
the concept of Lanczos
basis \cite{Lanczos}
and, in this spirit, we will assume that
precisely
such a basis is
used to span
the
$Q^{(+)}-$projected subspace of Hilbert space
with $Q^{(+)}=Q+D$.

Equivalently we will assume that
out of the model space, our sub-Hamiltonian
$Q^{(+)}\,H\,Q^{(+)}$
is a tridiagonal matrix so that we have
 \be
 H=
  \left[ \begin {array}{cccc|ccccc}
  H_{11}&\ldots
 &H_{1M-1}&H_{1M}&0&0&\ldots&\ldots&0
   \\
 \vdots&
 &\vdots&\vdots&\vdots&\vdots &&
 &\vdots
    \\
 H_{M-11}&\ldots
 &H_{M-1M-1}&H_{M-1M}&0&0&\ldots&\ldots
 &0
    \\
 H_{M1}&\ldots
 &H_{MM-1}&a_0&b_0&0&\ldots
 &\ldots&0
   \\
 \hline
     0&\ldots
 &0&c_1&a_1&b_1&0
 &\ldots&0
   \\
   0&\ldots
 &0
 &0&c_2&a_2&b_2&\ddots
 &\vdots
   \\
 \vdots&&\vdots &\vdots&0
 &\ddots&\ddots&\ddots&0
   \\
 \vdots&&\vdots &\vdots&\vdots&\ddots&c_{K-1}&a_{K-1}&b_{K-1}
    \\
  0&\ldots&0&0&0&\ldots&0&c_{K}&a_{K}
    \\
 \end {array} \right]\,.
 \label{kitie}
 \ee
For an efficient
``direct problem''
construction of $H_{ef\!f}(E)$ from $H$,
the basic benefits of such a choice
were outlined in our preceding sections.
In the inverse-problem setting we will still use ansatz
(\ref{kitie}) emphasizing that
one has to distinguish between the
known matrix elements
(represented by
the upper-case symbols)
and the
lower-case matrix elements which
have to be reconstructed using the knowledge of $H_{ef\!f}(E)$
of Eq.~(\ref{finkit}).


The number of the lower-case
matrix elements in Eq.~(\ref{kitie})
is assumed finite
so that their reconstruction using $H_{ef\!f}(E)$
would require just
the evaluation
of
function ${\cal G}(E)$
at
a finite number of
some
real values of $E=E_{\alpha}$
with, say, $\alpha=0,1,\ldots,2K$.
The choice of these values may be
random.
In fact, they should not coincide with the
bound-state energies. Indeed,
whenever we know a physical
energy (sub)spectrum,
we can immediately
solve the related reconstruction (sub)problem by
writing down the related part of $H$
in a diagonal-(sub)matrix form.
Hence, only the reconstruction of the rest of $H$ (if any)
would remain nontrivial.

A consequence of the tridiagonality of
matrix  $Q^{(+)}\,H\,Q^{(+)}$
is that
even when $N \to \infty$,
it enables us to evaluate
the resolvents
in terms of
continued fractions (cf.~(\ref{[9h]})).
This defines the
effective Hamiltonian (\ref{finkit})
which only depends on the products
$\rho_k=b_kc_{k+1}$
of the off-diagonal matrix elements of $H$.

This observation should be kept in mind also
during the reconstruction of $H$.
All of the values of $c_{k+1}$
have to be treated as arbitrary parameters,
therefore.
In other words, one can
suppress such a residual ambiguity
by
rescaling
the basis. This yields
the following final form
of the Hamiltonian to be reconstructed,
 \be
 H=
  \left[ \begin {array}{ccc|c|ccccc}
  H_{11}&\ldots
 &H_{1M-1}&H_{1M}&0&\ldots&&\ldots&0
   \\
 \vdots&
 &\vdots&\vdots&\vdots&&&
 &\vdots
    \\
 H_{M-11}&\ldots
 &H_{M-1M-1}&H_{M-1M}&0&\ldots&&\ldots
 &0
    \\
 \hline
 H_{M1}&\ldots
 &H_{MM-1}&a_0&\rho_0&0&\ldots
 &\ldots&0
   \\
 \hline
     0&\ldots
 &0&1&a_1&\rho_1&0
 &\ldots&0
   \\
   0&\ldots
 &\ldots
 &0&1&a_2&\ddots&\ddots
 &\vdots
   \\
 \vdots&&&\vdots&0
 &1&\ddots&\rho_{K-2}&0
   \\
 \vdots&&&\vdots&\vdots&\ddots&\ddots&a_{K-1}&\rho_{K-1}
    \\
  0&\ldots&\ldots&0&0&\ldots&0&1&a_{K}
    \\
 \end {array} \right]\,.
 \label{fkitie}
 \ee
We only have to remember that
at all $\,k=0,1,\ldots,K-1$
we have to re-factorize
$\,\rho_{k}=b_k c_{k+1}\,$ in order to
recover Hamiltonian $H$ of Eq.~(\ref{kitie}).

\subsection{Final form of the recipe}

Our task consists now in the
reconstruction
of
submatrix
 \be
 {\cal S}_{}^{(K)}= Q^{(+)}\,H\,Q^{(+)}=
  \left[ \begin {array}{cccccc}
  a_0&\rho_0&0&\ldots
 &\ldots&0
   \\
     1&a_1&\rho_1&0
 &\ldots&0
   \\
   0&1&a_2&\rho_2&\ddots
 &\vdots
   \\
 0&\ddots
 &\ddots&\ddots&\ddots&0
   \\
 \vdots&\ddots&0&1&a_{K-1}&\rho_{K-1}
    \\
  0&\ldots&0&0&1&a_{K}
    \\
 \end {array} \right]\,
 \label{spodkit}
 \ee
of the Hamiltonian.
The solution will be obtained via our
picking up $2K+1$
different values of parameter $E=E_\alpha$, and
after the evaluation
of the related energy-dependent matrix elements ${\cal G}(E_\alpha)$
of the preselected effective Hamiltonian.
Then,
at an arbitrary $K=N-M\geq 0$,
the $2K+1$ unknown matrix elements
of ${\cal S}_{}^{(K)}$ will be defined
by Eq.~(\ref{fundame}), i.e.,
after abbreviation
 \be
 \label{inpopo}
 {\cal G}(E_\alpha)={G}_\alpha\,,\ \ \ \ \alpha=0,1,\ldots,2K\,,
 \ee
by the
coupled set
of nonlinear algebraic equations
 \be
 H_{MM}-E_\alpha
 +\left [H\,Q\,\frac{Q}{E_\alpha-Q\,H\,Q}\,Q\,H\right ]_{MM}
 =
 {G}_\alpha
 \,,\ \ \ \ \alpha=0,1,\ldots,2K\,.
 \label{urfund}
 \ee
Here, the set of unknown quantities $a_0, \rho_0, \ldots, a_K\,$
enters the
left-hand-side resolvent via its
continued-fraction definition~(\ref{[9h]}),
 \be
 \label{qwopo}
 {\cal G}(E)=a_0-E-b_0f_1(E)c_1\,\equiv\,1/f_0(E)\,.
 \ee
Thus, we can rewrite the final implicit definition (\ref{urfund}) of $H$
in a compact form
 \be
  1/f_0(E_\alpha)={G}_\alpha\,,
 \ \ \ \ \ \alpha=0,1,\ldots,2K\,.
 \label{[13]}
 \ee
In such a set of coupled nonlinear algebraic equations
the left-hand-side expression
(having the form of continued fraction
of Eq.~(\ref{cf}) extended up to $k=0$)
can be also expressed as a ratio of two polynomials.
This means that Eq.~(\ref{[13]})
can be finally interpreted and solved as a coupled set
of $\,2K+1$ polynomial equations
determining the $2K+1$ matrix elements of
$\, {\cal S}^{(K)}$.

\section{Examples\label{inpro}}

In applications, the solution of the set of equations (\ref{[13]})
has to be performed using a suitable symbolic manipulation
software.
At the first few integers $K$, nevertheless,
the calculation can be also performed by hand.

\subsection{The choice of $N=2$ and $M=1$\label{illu}}

At $M=1$ and $K=1$, equation~(\ref{kitie})
does not contain
any ``known'' (i.e., upper-case)
matrix elements
so that we have to reconstruct
the whole two-by-two matrix ``full-space''
Hamiltonian
 \be
 H^{(2)}=
  \left[ \begin {array}{cc}
  a&b\\
  c&d
  \ea
  \right ]\,.
  \label{notyet}
 \ee
A full information about the quantum system in question
may still be assumed to be provided by
a hypothetical experiment yielding a
pair of bound-state energies, say, $E^{(1)}=X$ and $E^{(2)}=Y$.
Obviously,
the reconstruction
is ambiguous even when
the candidate matrix (\ref{notyet}) is assumed real, and
even when
the ``input'' values of $X$ and $Y$ become
reinterpreted as the roots of a more or less arbitrary
``effective Hamiltonian'' scalar function ${\cal G}(E)$.

In such an extreme toy model
the input information
can be directly related
to the matrix elements of
the ``unknown'' but ``exact'' two-by-two Hamiltonian (\ref{notyet}).
Due to the validity of the standard secular equation
the connection has the form of
the following coupled pair of algebraic
equations,
 \be
 X^2-({a}+d)\,X + {a}d-bc=0\,,\ \ \ \
 Y^2-({a}+d)\,Y + {a}d-bc=0\,.
 \label{tarof}
 \ee
Their inspection
reveals that two
out of the four
unknown quantities ${a}$, $b$, $c$ and $d$ only appear here in
the form of product $bc=\varrho$. Thus,
either $b$ or $c$
has to be interpreted as a free
parameter.
This is a freedom reflecting
the unitary-transformation
variability of the basis. The sought
matrix (\ref{notyet})
can be replaced, therefore,
by its simpler
three-parametric isospectral
avatar (\ref{fkitie}),
 \be
 H^{(2)}=
  \left[ \begin {array}{cc}
  {a}&\varrho\\
  1&d
  \ea
  \right ]\,.
  \label{bnotyet}
 \ee
Incidentally, there exists another,
dynamical-input ambiguity
connected with our freedom of
the choice of the origin on the real line of the energies.
Thus, once we choose this origin as equal to the
mean value of the
(say, real and non-degenerate)
``input'' two-level spectrum, i.e., when we
set $X + Y =0$, the subtraction of Eqs.~(\ref{tarof})
enables us to eliminate parameter $d$,
 $$
 d=d({a})=-{a}\,.
 $$
The remaining equation defines $\varrho$,
 \be
 \varrho=\varrho({a})=X^2-{a}^2.
 \label{thekey}
 \ee
The reconstruction of $H^{(2)}$ of Eq.~(\ref{bnotyet})
based on the single physical input parameter $X=-Y$
or on an arbitrary
measurement-fitting scalar function
${\cal G}(E) $
is completed.

The value of matrix element ${a}$
remains unconstrained.
We are free to prefer its ``small-value'' choice such that
${a}^2 < X^2$. This
would enable us to
set $b=c$ and get the ultimate Hamiltonian
in its most common Hermitian {\it alias\,}
symmetric-matrix form (\ref{notyet}).

\subsection{Closed-form solution of the $K=1$ inverse problem at any $M$}

As long as we decide to keep all of
our matrix dimensions $M$, $K$ and $N=M+K$ finite,
the left-hand-side expression
in (\ref{[13]})
is
expressible as the ratio of polynomials
in all of the elements of ${\cal S}_{}^{(K)}$.
At the first sight, the
task of solving such a coupled set of nonlinear
algebraic equations does not
look feasible, indeed.

The skepticism was suppressed
after we succeeded in
solving this system of algebraic equations at
$K=1$. In this case one has
to determine
the three matrix elements in
 \be
 {\cal S}_{}^{(1)}=
  \left[ \begin {array}{cc}
  a_0&\rho_0
   \\
     1&a_1
      \\
 \end {array} \right]\,.
 \label{1spodkit}
 \ee
The triplet of equations
(\ref{[13]}) can be then given
the coupled-polynomial algebraic form
 \be
 {E_{\alpha}}^{2}+G_{\alpha} \, E_{\alpha}
  -  E_{\alpha}\,\left( a_{{0}}+a_{{1}} \right)
  -G_{\alpha}\,a_{{1}}
  +a_{{0}}
 a_{{1}}-{\it \rho}_{{0}}=0\,,\ \ \ \ \alpha=0,1,2\,.
 \label{ome}
 \ee
These equations are linear in the three
new unknown parameters defined as follows,
 \be
  x_{{1}}=-a_{{0}}-a_{{1}}\,,\  \ \
  y_{{1}}=a_{{1}}\,,
  \ \ \
  x_{{2}}=a_{{0}}a_{{1}}-{\it \rho}_{{0}}
 \,.
 \label{toma}
 \ee
This is a change of variables which converts Eqs.~(\ref{ome})
into the following linear algebraic
matrix-inversion problem
 \be
 \ba
 G_{0}\,y_{{1}}
 -E_{0}\,x_{{1}}
 -x_{{2}}=
 {E_{0}}^{2}+G_{0}  E_{0}\,,\\
 G_{1}\,y_{{1}}
 -E_{1}\,x_{{1}}
 -x_{{2}}=
 {E_{1}}^{2}+G_{1}  E_{1}\,,\\
 G_{2}\,y_{{1}}
 -E_{2}\,x_{{1}}
 -x_{{2}}=
 {E_{2}}^{2}+G_{2}  E_{2}\,.
 \ea
 \label{epitoma}
 \ee
This is
a routinely tractable task
which only converts the
``experimental''
information about dynamics
as carried by the
triplet of choices
of energies $E_{\alpha}$ and
of the related values $G_{\alpha}$ of functions (\ref{inpopo})
with $\alpha=0,1,2$
into a ``compressed''
information about dynamics
as carried by $x_1$, $x_2$ and $y_1$.

This was the first, trivial step of the two-step reconstruction
of the full-space Hamiltonian (\ref{fkitie})
or,  at
$K=1$ and any $M$, of its unknown submatrix (\ref{1spodkit}).
For a completion of the task
it
would be now sufficient to invert, in the second step, the
nonlinear mapping
 $\{a_0,a_1,\rho_0\} \to \{x_1,x_2,y_1\}$
as defined by Eq.~(\ref{toma}).
Fortunately, as long as $K=1$, the answer is easy:
 \be
  a_{{1}}=y_{{1}}\,,\ \ \ \
 a_{{0}}=-x_{{1}}-y_{{1}}\,,\ \ \ \
 {\it \rho}_{{0}}=-x_{{1}}y_{{1}}-x_{{2}}-{y_{{1}}}^{2} \,.
 \label{epi}
 \ee
The ultimate inversion of the mapping of Eq.~(\ref{toma})
is
given by closed, non-numerical formulae.

\section{Discussion\label{discussion}}

\subsection{A note on benefits in quantum physics}

In contrast to the Feshbach's reduction $H \to H_{eff}(E)$
(which requires just a
trivial insertion in formula (\ref{formu})),
the inverse problem  $H_{eff}(E) \to H$
is even, as we saw, difficult to define. 
At the same time, a reasonably persuasive
formulation of the inverse problem would certainly be perceived 
as a badly needed 
step of progress in physics.

Unfortunately, no sufficiently satisfactory 
and exhaustive formulations
of the Feshbach-inspired inverse problem
seem to be available in the market at present.
This is just a fact, with some of the consequences 
explained above. In other words, one could 
characterize, 
as unsatisfactory, even
our present formulation of the inverse problem itself,
with the weaknesses ranging
from our set of assumptions (which could be perceived
as over-restrictive) up to our final algebraic $K=1$
result (\ref{epitoma}) + (\ref{epi}) -- with
any extension to be expected more
complicated, less user-friendly and, 
last but not least, hardly suitable
for a letter-journal.

So why to still bother about the inverse problem? 
What are the expected benefits?
The answers are encouraging and reflecting, paradoxically,
a truly enormous success
of the Feshbach's idea of construction  $H \to H_{eff}(E)$.
A hidden reason is that this success is also
responsible for the emergence of a deeply erroneous
persuasion that 
even if we do not know $H$, we 
do not need to feel bothered.
We
may feel free to
make a more or less arbitrary 
guess of a user-friendly $ H_{eff}(E)$
and, subsequently, we 
may try to persuade the experimentalists
(who decide to buy our guess)
that the corresponding
exact and energy-independent $H$ does exist, and that it
would only be too complicated
to be worth a reconstruction.

Even on a purely intuitive level
such a model-building philosophy could be misleading.
In fact, our present project was motivated precisely
by our equally intuitive persuasion that
in applications, 
the vast majority of the descriptions of reality
based on the user-friendly $ H_{eff}(E)$s remains 
deeply inconsistent. By {\em not\,} leading,
after any hypothetical reconstruction, to {\em any\,}
energy-independent
full-space Hamiltonian $H$ at all. 

In the present paper we made the first step towards
a less intuitive argumentation. 
We accepted a number of technical assumptions
concerning the ``unknown'' $H$
which, by themselves,
restricted also the class of the ``eligible''
initial  $ H_{eff}(E)$s. 
This gave us a (not entirely expected)
chance of being constructive, and also an equally unexpected
chance of being concise, explicit and non-numerical.
With these features forming a
rather abstract methodical benefit of our present partial 
(i.e., still strongly assumption-dependent)
resolution of the Feshbach-inspired inverse problem.

This being achieved, let us now add
a few more specific applicability-related remarks.
The first one concerns the freedom, or benefit, of having our 
model-space dimension $M$ 
(say, in Eq.~(\ref{enproj}))
arbitrarily large.
This already is an advantage {\it par excellence\,}
because with $M \gg 1$ we usually 
reach a reasonable precision using even
the effective model (\ref{ethee}) with trivial $K=0$.
In this setting, even our ``first step'' $K=1$ inversion and
formulae would already offer a good and constructive test of 
the expected weakness of the $M-$dependence of the results.

The experienced users of the theory could immediately object that
in the latter $M \gg1$ extreme also
the relevance of any choice of 
${\cal G}(E)$ becomes suppressed.
In this sense, the  
arguments {\it pro et contra\,}
should be tested at the smaller $M$.
In spite of the deeply conditional
nature of any model with a small $M$,
even their study using reconstruction
may
throw new light on the benefits and/or 
paradoxes of the applied inversion
$H \to H_{eff}(E)$.

For the purposes of illustration let us
consider another schematic toy model with $M=2$ and $K=1$.
For the reconstruction
 \be
 H^{(2)}_{eff}(E)=
  \left[ \begin {array}{cc}
  A&B\\
  C&{\cal G}(E)+E
  \ea
  \right ]\,
  \ \ \ \ \to
  \ \ \ \  
 H^{(3)}=
  \left[ \begin {array}{ccc}
  A&B&0\\
  C&a_0&\rho_0\\
  0&1&a_1
  \ea
  \right ]\,
  \label{ytyet}
 \ee
of the ``full-space'' three-by-three matrix
Hamiltonian
we 
are given here just
the information that
at some two energy levels $E_\pm$
provided by the experimentalists we have
 \be
 \det \left (H^{(2)}_{eff}(E)-E
 \right )=0
 \label{taroz}
 \ee
Using our above-outlined formalism
we evaluate the ``missing'' matrix elements
$a_0$, $\rho_0$ and $a_1$ of the reconstructed Hamiltonian.
Alas, what
we get will be the result which will strongly depend
on our {\em 
guess of the function\,} ${\cal G}_{input}(E)$
at $E \neq E_\pm$.

A key point is that the latter guess 
(interpolating between 
the only available
``experimental input'' 
$E_-$ and $E_+$)
is still 
strongly ambiguous.
In principle, therefore,
we may get a ``satisfactory'' result
(which means that we find the three eigenvalues of $H^{(3)}$,
with the new, third one confirmed, say, by the experimentalists),
or not.
This is a paradox: The reconstruction appears successful or
ceases to be acceptable in 
a way which is not under our control.

The essence of the explanation is straightforward,
in our illustration using $M=2$ at least.
One simply notices that
the essence of 
the difference between the
``lucky'' and ``wrong'' guesses of ${\cal G}_{input}(E)$
is that among all of them,  
there is one which is unique and optimal,
given, in closed form, by Eq.~(\ref{taroz}),
 \be
 {\cal G}_{(M=2)}(E)=BC/(A-E)\,.
 \label{sumava}
 \ee
Thus, all of our $M=2$ ``input guesses''
are found classified 
by the degree of their coincidence with the ``universal''
function (\ref{sumava}).

For this reason, a desirable enhancement of such a coincidence
may be expected to require the choice of a larger $M$ in practice.

\subsection{Broader context}

In spite of a comparatively elementary nature of our two above-mentioned
illustrative $K=1$ examples,
their deeper inspection
indicates
that
in general,
the sign of the product $\rho_0=b_0c_{1} $
can happen to be negative.
This opens an entirely unexpected new set of methodical questions.
Indeed, such a value could result from an innocent-looking
choice of tentative
``experimental-input'' values
$x_1=0$, $x_2=-3$ and $y_1=2$ yielding
$a_0=-2$, $a_1=2$ and negative $\rho_0=-1$.

In such a situation
the reconstructed
full-space energy-independent Hamiltonian $H$
becomes manifestly
non-Hermitian.
In the context of quantum mechanics
this means that having the experiment-provided
spectrum which is real,
such a Hamiltonian
would necessarily be
quasi-Hermitian
and Hermitizable
(cf., e.g., \cite{Geyer,book} for
the theory and
explanation).

The latter remark can be read as an explicit demonstration that
our present, mathematically motivated return to
the old Feshbach's inverse problem
might bring new benefits also in
the quickly developing field of non-Hermitian physics
(cf., e.g.,
some its most recent reviews \cite{Christodoulides,Carlbook,Mousse}).
In this context, indeed,
we see that a certain new form of unification
connecting the non-Hermitian
and Hermitian physics
emerges
as one
of the
consequences of the reconstruction
of ``fundamental theory'' (i.e., of the
energy-independent operator $H^{(N)}$)
from some of its experimental aspects (i.e., from the measured
data
encoded, in our present notation,
in the scalar function ${\cal G}(E)$ of the real
energy parameter).

The emergence of the ``anomalous''
non-Hermitian Hamiltonians and/or effective Hamiltonians
(as recently reviewed, e.g., in  \cite{Christodoulides,Carlbook})
can even be traced back to
the classical Feshbach's
paper \cite{Feshbach}.
Up to these days, therefore, the
subject still re-emerges as topical
in nuclear physics.
Even
certain low-lying states are not fully reproduced there
by the currently available model-space Hamiltonians.
In \cite{nuclphys}, for example, an inclusion of
a missing proton-neutron term
the nuclear effective interaction has been found
crucial for astrophysical applications.

Needless to add here that there is no chance to cover or sample the
multiple branches of the related current
effective-interaction-related
research.
At random,
let us just mention the very recent study
\cite{molecular} in atomic physics
(fitting the rotational and hyperfine-structure spectrum
of a certain bi-alkalic Na+Cs molecule)
or another equally recent contribution
\cite{cinani} to nuclear physics
amending the shell-model description of the nuclei of several isotopes of neon:
Notably, the latter paper highlighted the shell closure at ${\cal N}=14$
but reported a remarkably low excitation energies of $^{30}Ne$
suggesting that the conventional magical number status of ${\cal N}=20$
might be put under question-mark.

\section{Summary\label{summary}}

In virtually all of the existing
applications of the effective Hamiltonians
a weakening of their energy-dependence
belonged to
the mainstream of studies of their amendments
(see, e.g., a review paper
\cite{recurs} in this context).
For this reason we believe that
our present approach to this problem based on the
solution of coupled polynomial algebraic equations
has a capacity of throwing new light on these
efforts and developments.

In our text we presented
a constructive analysis of
the mechanism of a
systematic weakening of the energy-dependence
of
$H_{ef\!f}(E)=
H_{ef\!f}^{(M)}(E)$
with the growth of
the model-space dimension $M$.
The overall method
appeared constructive and straightforward. Its first step
(consisting in a rearrangement of the input-data information about dynamics)
appeared easy (requiring just an inversion of a matrix).
Still, the essence of the reconstruction of $H$
appeared nontrivial, requiring an
explicit solution of a coupled set of polynomial algebraic
equations as sampled here, at $K=1$, by Eq.~(\ref{epi}).

In a final remark we should add that
at any $K$,
the process of reconstruction of $H$ can lead to
the emergence of
products $\rho_k=b_kc_{k+1} $
with a
negative value at one or more subscripts $k$.
This is an apparent anomaly which was proved to exist
even at $K=1$.
Still, the same explanation
as given towards the end of subsection \ref{illu}
remains applicable at any $K$:
The ``anomalous'', non-Hermitian
reconstructed Hamiltonians $H$ remain Hermitizable
and, therefore, fully compatible with the standard
postulates of quantum theory.

\end{document}